\documentclass[twocolumn,prb,color,superscriptaddress,psfig,showpacs,amsmath,amssymb]{revtex4}

\usepackage{epsf}

\usepackage{natbib}
\setcitestyle{square,numbers}

\usepackage{tabularx} 
\usepackage{graphicx} 
\usepackage{epstopdf}
\usepackage{epsf}
\usepackage{dcolumn}
\usepackage{bm}
\usepackage{color, xcolor}

\usepackage{multirow}

\usepackage[normalem]{ulem}

\usepackage{color,soul}
\setul{0.5ex}{0.5ex}
\setulcolor{red}

\begin{document}

\title{Altermagnetism in 6H perovskites}

\author{Sergey V. Streltsov}
\affiliation{Institute of Metal Physics, Ural Branch of the Russian Academy of Sciences, Ekaterinburg 620137, Russia}
\affiliation{Department of theoretical physics and applied mathematics, Ural Federal University, Ekaterinburg 620002, Russia}

\author{Sang-Wook Cheong}
\affiliation{Keck Center for Quantum Magnetism and Department of Physics and Astronomy, Rutgers University, Piscataway, NJ 08854, USA}

\date{\today}

\begin{abstract}
The combination of a centrosymmetric crystallographic structure with local structural alternations and collinear antiferromagnetism can lead to broken PT (Parity $\times$ Time-reversal) symmetry, resulting in altermagnets with non-relativistic spin-split bands. The 6H perovskites with composition A$_3$BB'$_2$O$_9$ exhibit unique layered structural alternations and typically adopt an antiferromagnetic ground state.  Here, we report the discovery that several 6H perovskites are indeed altermagnets exhibiting non-relativistic spin-split bands. We also explore the possible presence of net magnetization due to spin-orbit coupling in these materials, as well as the manifestation of giant piezomagnetism.  Since the single crystals of 6H perovskites can be readily grown and cleavable, our findings provide a new avenue to study the cleaved atomically-flat surfaces of altermagnets with advanced experimental techniques such as spin-resolved scanning tunneling microscopy (STM) or spin-resolved angle-resolved photoemission spectroscopy (ARPES) to explore their spin splitting nature.
\end{abstract}

\maketitle

\hfill \break {\bf \noindent INTRODUCTION\\} 

Perovskites are one of the most studied structural classes, encompassing not only the popular perovskite solar cells and the famous cuprate superconductors but also many other materials important for various applications. In so-called cubic perovskites, such as SrTiO$_3$ or LaMnO$_3$, transition metal ions are surrounded by ligand's octahedra, forming a cubic or distorted cubic lattice where the octahedra share corners. Depending on the ionic radii of the A and B elements in the general formula ABX$_3$, other types of packing can be realized\cite{de2024}. In hexagonal perovskites, such as BaNiO$_3$, AX$_3$ layers form a hexagonal close-packed structure, and the BX$_6$ octahedra share faces rather than corners.

Both cubic and hexagonal perovskites demonstrate extraordinary physical properties such as magnetoresistance~\cite{kobayashi1998}, high dielectric constant~\cite{merz1949}, unusual magnetic, charge and orbital orders~\cite{streltsov2017},  many of them were found to be not only ferroelectrics, but also multiferroics~\cite{cheong2007,khomskii2009}, other perovskites turned out half-metals with record high Curie temperatures~\cite{liu2022}. 

Very recently a new notion of altermagnetism was introduced to characterize materials, which exhibit ferromagnet behavior, but have zero net magnetization~\cite{noda2016,hayami2019,naka2019,ahn2019,yuan2020a,smejkal2022}. This class of magnetic materials demonstrate anomalous electronic and spin transport, Nernst and magneto-optical effects etc.~\cite{smejkal2022,guo2023,song2025}  Different perovskites have been suggested to be altermagnets: CaCrO$_3$~\cite{naka2021,streltsov2008}, LaTiO$_3$~\cite{maznichenko2024}, HgMnO$_3$~\cite{myakotnikov2024}, (La,Ca)MnO$_3$~\cite{vsmejkal2020,solovyev1997} and many others. 
\begin{figure}[b!]
\includegraphics[width=0.46\textwidth]{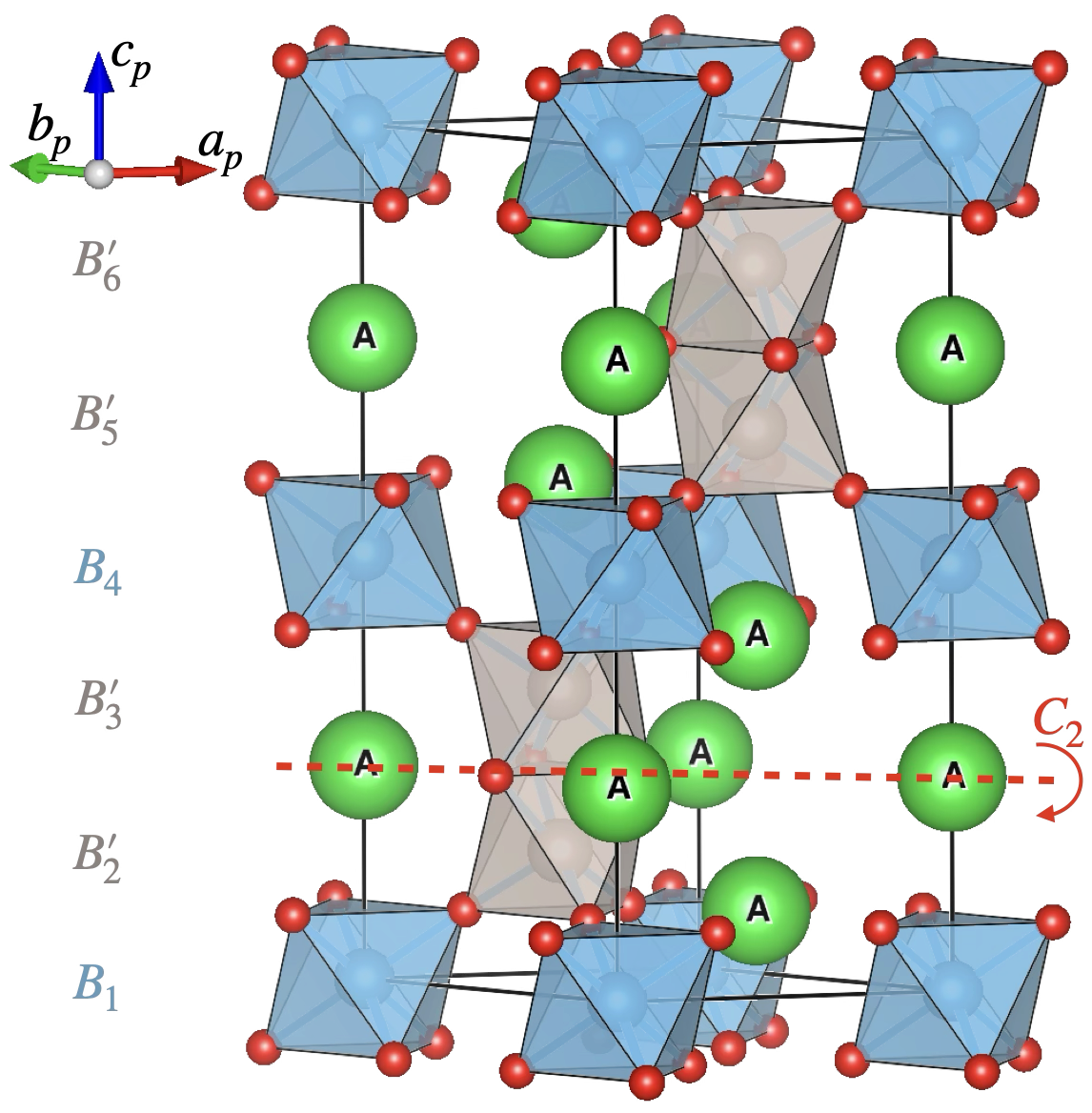}
\caption{\label{crystal-structure} The crystal structure of 6H perovskites. Magnetic ions can occupy $B$ positions  to form triangular lattice and $B'$ constituting dimers, indexes numerate layers. For clarity a primitive unit cell (which is two times smaller than the crystallographic unit cell) is shown. The primitive $c_p$-axis coincides with the conventional $c$-axis. One of the $C_2$ rotation axes (at $c/4$) is sketched.  }
\end{figure}

However, the simplest ABX$_3$ perovskites are a tip of iceberg and progress in chemistry resulted in synthesis of mixed structures, when layers of cubic perovskites are intertwined with one or few layers of hexagonal (H) perovskites. In this paper we discuss altermagnetism in these mixed perovskites, so-called 6H perovskites~\cite{hinatsu2015a,nguyen2021a,de2024} with general formula A$_3$BB'$_2$O$_9$ and the unit cell (u.c.) consisting of 6 repeating layers as shown in Fig.~\ref{crystal-structure}. Magnetic ions can occupy $B$ and $B'$ positions, both having octahedral coordination. Two B'O$_6$ octahedra share their faces and form B'$_2$O$_9$ dimers. Both dimers and ``isolated'' B ions form their own triangular lattices. While the magnetic structure  for most of 6H perovskites remains unsolved we argue that many of them can be altermagnets. Detailed {\it ab initio} calculations for two of them show not only corresponding spin-split band structure, various magneto-optical effects, but also a giant piezomagnetic response, which makes a whole class of 6H perovskite interesting for further experimental and theoretical studies.

\hfill \break {\bf \noindent SYMMETRY ANALYZES\\} 

\begin{table*}[t!]
 \begin{tabular}{lcccccccc}  
  \hline \hline
Compound & Magnetic & $T_N$ & Magnetic     & Magnetic  & Altermagnetism & Hall &  Reference\\
         &ions      &       & point group  & structure & type           &      &                &\\
 \hline
Ba$_3$CoIr$_2$O$_9$  & Co$^{2+} (d^7)$, Ir$^{5+}(d^4)$  & 107 K& $2/m$ & $\approx || c$ & M-type & $\rho_{xz}$ 
& \cite{garg2021a,guo2023}\\
Ba$_3$SrIr$_2$O$_9^*$ & Ir$^{5+} (d^4)$ & - & $2'/m'$ & coll. in $ac$ & M-type & $\rho_{xy},\rho_{yz}$  & \cite{nag2018}\\  
\hline
Ba$_3$NiRu$_2$O$_9$  & Ni$^{2+} (d^8)$, Ru$^{5+} (d^3)$  & 5 K & $6'/m'mm'$ & $||c$ & S-type & no  & \cite{Lightfoot1990,guo2023,supp}\\
Ba$_3$TbRu$_2$O$_9$  & Tb$^{4+} (f^7)$, Ru$^{4+} (d^4)$  & 9.5 K & $6'/m'mm'$ & $||c$ & S-type & no & \cite{doi2001a,doi2002a}\\
 \hline \hline
  \end{tabular} 
\caption{\label{6H-AM-table} Altermagnets with 6H perovskite structure. Classification of altermagnetism type is according to \cite{cheong}. For Ba$_3$SrIr$_2$O$_9$ the theoretical magnetic ground state obtained in previous DFT+U+SOC calculations were used for analysis; moments are collinear in the $ac$ plane with large ferromagnetic component.}
\end{table*}
\begin{figure}[b!]
\includegraphics[width=0.49\textwidth]{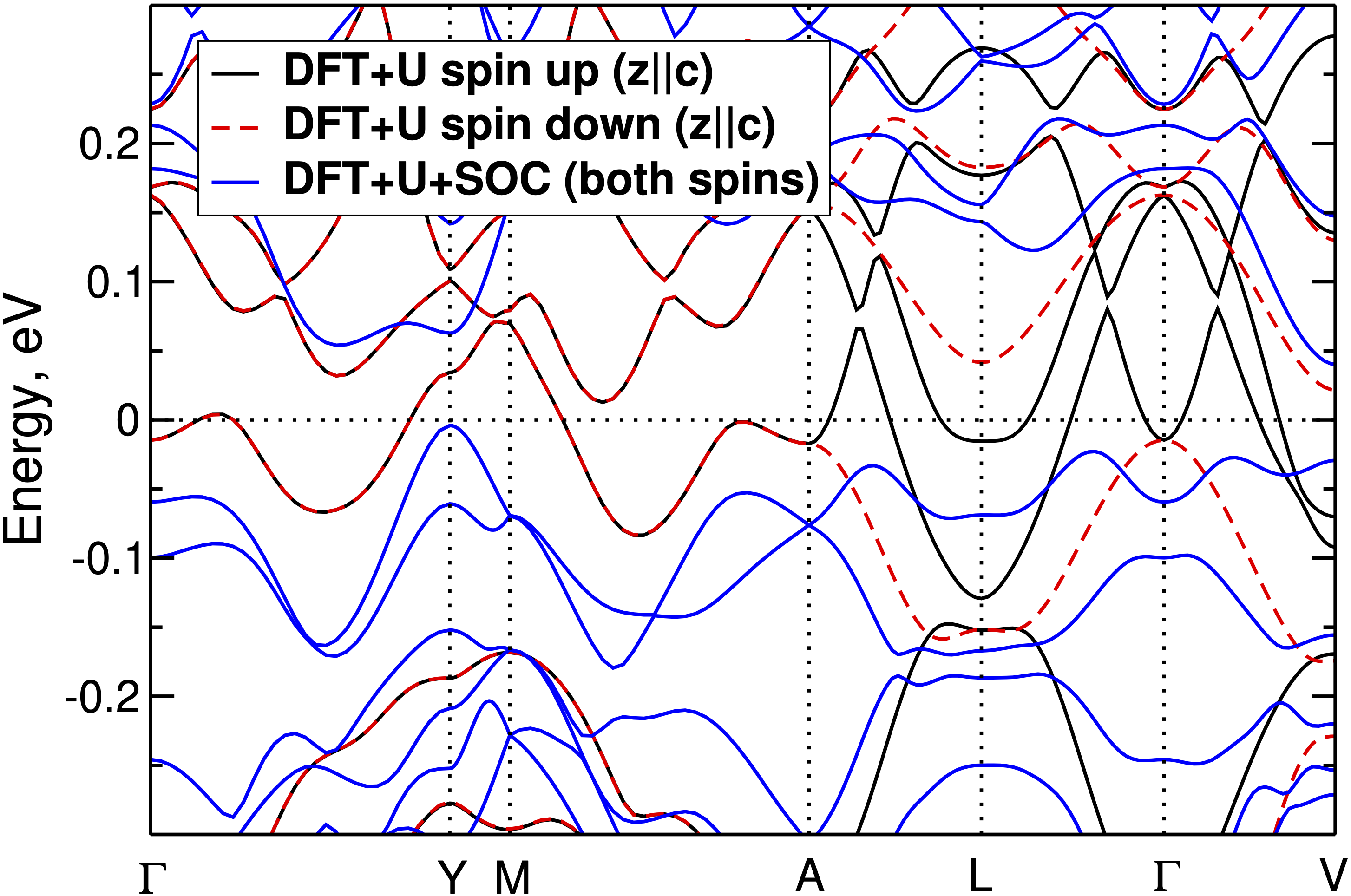}
\caption{\label{CoIr-bands} The band structure of Ba$_{3}$CoIr$_2$O$_9$ obtained using different approximations is shown. Clear evidence for altermagnetism is observed in DFT+U calculations when moving along the $L$-$\Gamma$-$V$ direction (black and red lines); however, the material remains metallic in this approach. Including spin-orbit coupling (SOC) and a non-collinear magnetic structure leads to an insulating state (blue lines). Since SOC mixes spin up and down states, both of them are plotted for DFT+U+SOC. The Fermi energy is set to zero. High-symmetry points of the Brillouin zone are shown in Fig.~\ref{Ba3CoIr2O9-FS}.
DFT+U+SOC band structure weighted according to different spin projections is also presented in Fig. S2 of Supplemental Materials.}
\end{figure}

There are three possible structures characterized by the hexagonal $P6_3/mmc$, orthorhombic $Cmcm$, and monoclinic $C2/c$ space groups, which describe A$_3$BB'$_2$O$_9$  perovskites with $B'$  ions forming dimers (the hexagonal $P6_3mc$, where only half of the sites in dimers are occupied by transition metals, and the monoclinic $P2_1/c$  structures without dimers are not considered in this study)~\cite{de2024}. There is no inversion center ($I$) connecting the magnetic $B$ sites, but $C_2$ rotation axes perpendicular to the $c$-axis (located at 1/4 and 3/4 of $c$ for $Cmcm$ and $C2/c$, and additionally at the origin and $c/2$ for $P6_3/mmc$). This makes 6H perovskites altermagnetic if the two $ab$ planes of magnetic $B$ ions (at the origin and $c/2$) are antiferromagnetically ordered and the spins are (mostly) aligned along the $c$-axis. The same $C_2$ rotation rather than inversion transforms $B'$ sites of {\it the same} dimer into each other.

While there are many $6H$ perovskites with magnetic ions occupying $B$ or $B'$ sites, and many of them show clear signatures of dominating antiferromagnetic (AFM) interactions, a detailed study of the ground-state magnetic structure has been performed for only 15 AFM 6H perovskites. Detailed lists of ferromagnetic, ferrimagnetic, antiferromagnetic $6H$ perovskites, as well as those that do not exhibit long-range magnetic order being potential candidates for quantum spin-liquid behaviour or showing thermal induced magnetization, are presented in Tables S1-S3 of the supplemental materials~\cite{supp}.

Symmetry analysis by magnetic point groups (MPG) according to \cite{cheong} shows that 4 out of 15 AFM 6H perovskites appear to be altermagnets. Their properties are summarized in Tab.~\ref{6H-AM-table}. There are two M-type altermagnets, Ba$_3$CoIr$_2$O$_9$ and Ba$_3$SrIr$_2$O$_9$, with broken time-reversal symmetry, where one would expect non-vanishing magnetization due to orbital magnetism (spin-orbit coupling). In addition, time-reversal symmetry is broken in Ba$_3$NiRu$_2$O$_9$ and Ba$_3$TbRu$_2$O$_9$, which must be S-type altermagnets with symmetric non-relativistic spin splitting. Next, we use density functional theory (DFT) calculations to verify these conclusions and study the implications of altermagnetism in one representative from each class.

\hfill \break {\bf \noindent DFT RESULTS\\}
\begin{figure}[b!]
\includegraphics[width=0.3\textwidth]{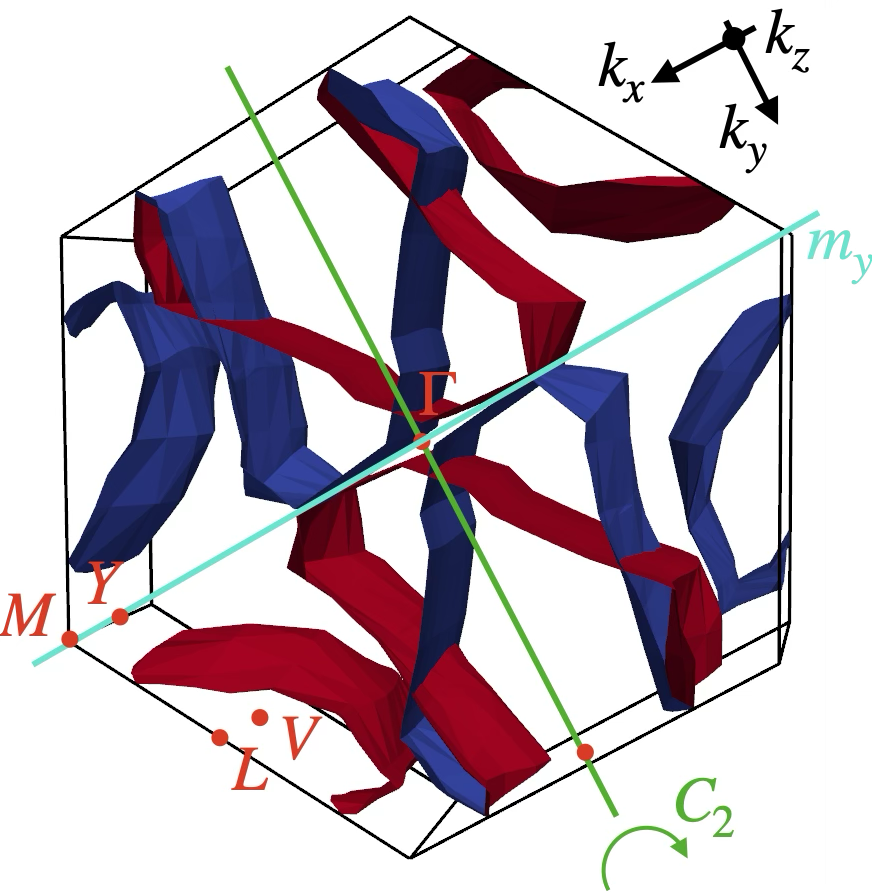}
\caption{\label{Ba3CoIr2O9-FS} The Fermi surface obtained in DFT+U calculations for Ba$_3$CoIr$_2$O$_9$ is shown. Different spin projections are indicated by blue and red. The $C_2$ rotation transforms all spin-up states into spin-down sheets. High-symmetry points of the Brillouin zone are marked in red (the $A$-point, located above the $\Gamma$-point, is not shown). $m_y$ is the mirror plane.}
\end{figure}

{\bf M-type altermagnet: Ba$_3$CoIr$_2$O$_9$}.
In M-type altermagnets, one would expect a spontaneous Hall effect if they were metallic. However, a general feature of stoichiometric 6H perovskites is that they are typically insulators. Transition metal ions occupying $B$ sites (in the case of Ba$_3$CoIr$_2$O$_9$, these are Co$^{2+}$, $3d^7$) are far apart and do not share common ligands. This results in small bandwidths, which can hardly compete with a substantial Hubbard $U$. The $B'$ sites (Ir$^{5+}$, $5d^4$) form dimers with corresponding molecular orbitals, and some electrons are delocalized over the dimer~\cite{streltsov2017}. However, since the dimers are also isolated, this does not lead to metallic conductivity.

However, our situation is even more complicated. Ba$_3$CoIr$_2$O$_9$ turns out to be metallic in DFT+U calculations, as shown in Fig.~\ref{CoIr-bands}. The band structure clearly demonstrates strong non-relativistic spin-splitting along the $L$-$\Gamma$-$V$ path, confirming the symmetry analysis. Indeed, the points $Y$, $M$, and $A$ all belong to the $m_y$ mirror plane (which is perpendicular to $k_y$). Therefore, these points are transformed only by the $C_2$ rotation (about the $k_y$ axis) and inversion $I$. If we first apply the $C_2$ rotation, we change $k_x \to -k_x$, $k_z \to -k_z$, and $S_z \to -S_z$. Then, the consecutive application of $I$ transforms these points back to their initial positions but with a flipped spin (since inversion does not affect the spin). Thus, there must be spin degeneracy: $\varepsilon(\vec{k} \uparrow) = \varepsilon(\vec{k} \downarrow)$ for any $k$-point belonging to the $m_y$ mirror plane, including $Y$, $M$, and $A$. There is no such restriction for the $L$ and $V$ points ($m_y$ is active for them and additionally flips the spin, leading only to the trivial condition $\varepsilon(\vec{k} \uparrow) = \varepsilon(\vec{k} \uparrow)$).

The altermagnetism also manifests itself in the Fermi surface (see Fig.~\ref{Ba3CoIr2O9-FS}), where sheets corresponding to different spins are transformed into each other by the $C_{2}$ rotation discussed in the previous section. 
\begin{figure}[b!]
\includegraphics[width=0.49\textwidth]{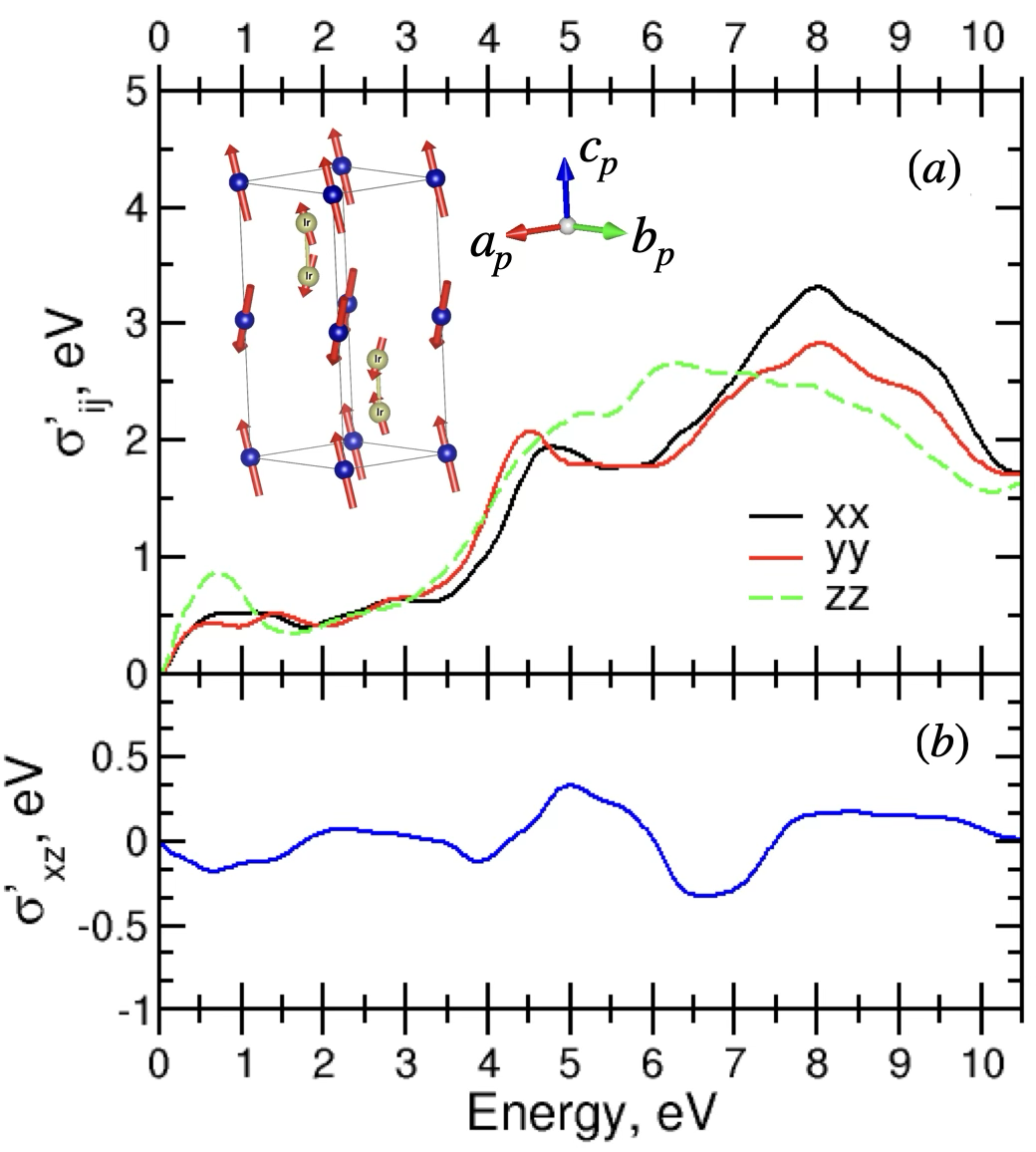}
\caption{\label{Ba3CoIr2O9-sigma}  The real part of the optical conductivity for Ba$_{3}$CoIr$_2$O$_9$, as obtained in DFT+U+SOC calculations, is shown. (a) shows three (symmetry independent) diagonal components, while (b) presents its single non-zero off-diagonal element, which serves as direct evidence of a magneto-optical effect. The magnetic configuration obtained in DFT+U+SOC calculations is presented in inset (Co is in blue, Ir is in grey), index $p$ indicates that the primitive cell is shown.
}
\end{figure}

The spin-orbit coupling strongly changes the electronic structure of Ba$_3$CoIr$_2$O$_9$ and leads to the opening of the 40 meV band gap (blue lines in Fig.~\ref{CoIr-bands})~\cite{garg2021a}. Therefore, Ba$_3$CoIr$_2$O$_9$ can be classified as a spin-orbit-assisted Mott insulator~\cite{streltsov2017} with zero Hall conductivity. Note, that even if  $U_{\rm Co}$ and $U_{\rm Ir}$ are increased up to unrealistic 10 eV and 3 eV, respectively, this material remains metallic in DFT+U. The spin-orbit coupling is essential for the insulating ground state. The magnetic symmetry should affect not only static properties but also frequency-dependent (optical) conductivity, $\sigma_{ij}(\omega)$, which is more relevant for insulators. For the $2/m$ MPG, there can be a single off-diagonal matrix element, $\sigma_{xz}(\omega)$ (and $\sigma_{zx} = - \sigma_{xz}$). Results of direct $\sigma_{ij}(\omega)$ calculations using the Green-Kubo formalism~\cite{gajdovs2006} within the DFT+U+SOC approach for the experimental magnetic structure, presented in Fig.~\ref{Ba3CoIr2O9-sigma}, confirm that this is indeed the case.

 One can probe the same physics by the magneto-optic Kerr effect (MOKE) experiments. Symmetry consideration shows that there must be 8 non-zero elements in $q_{ijk}$ tensor relating components of the magnetic field $H_{k}$ with dielectric permeability tensor $\beta_{ij} = q_{ijk} H_{k}$: 
$q_{xxy}$, $q_{yyy}$, $q_{zzy}$, $q_{yzx}$, $q_{yzz}$, $q_{xzy}$, $q_{xyx}$, $q_{xyz}$.

\begin{figure*}
    \centering
\includegraphics[width=\textwidth]{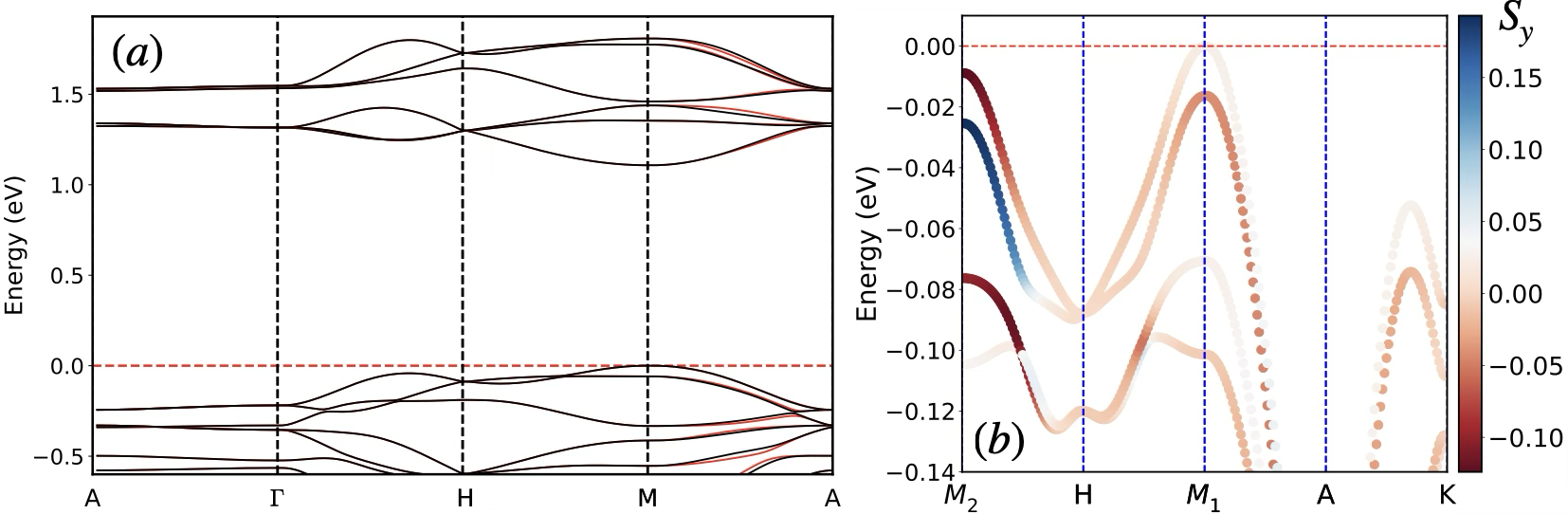}
\caption{\label{Ba3NiRu2O9-ek} Band structure of Ba$_3$NiRu$_2$O$_9$ obtained in (a) collinear AFM DFT+U calculations for experimental crystal structure (different spin projections are shown by color) and (b) in non-collinear DFT+U+SOC calculations with 1\% tensile strain applied along $y$ direction. In (b) color is used to show $S_y$ of corresponding state. Strain makes $M$-points inequivalent. $M_1$, $M_2$, and $K$ points are shown in inset of Fig.~\ref{MvsDelta}(a), while $A$ and $H$ are located above $\Gamma$ and $K$ points, respectively. The Fermi energy is in zero in both plots.}
\end{figure*}

The M-type of altermagnetism implies weak ferromagnetism via the canting of magnetic moments. The presence of heavy Ir$^{5+}$ ions with a large spin-orbit coupling constant and Co$^{2+}$ ions, which are prone to having large orbital moments, supports this tendency. This is exactly what is observed in Ba$_3$CoIr$_2$O$_9$ in DFT+U+SOC calculations. The resulting magnetic structure, respecting the $C_2$ rotation axes, is shown in the inset of Fig.~\ref{Ba3CoIr2O9-sigma}(a). A net ferromagnetic moment of 1.4 $\mu_B$ per formula unit (f.u.) is directed approximately along $y$. The canting of magnetic moments exceeds $20^{\circ}$ on Co and $25^{\circ}$ on Ir, which has a larger spin-orbit coupling. The $x$ and $y$ components of magnetic moments on different Co atoms are nearly the same by absolute values, but $x$ component alternates (according to $2/m$ MPG) and does not contribute to the net magnetization. Situation with Ir moments is the same. Isolated Ir$^{5+}$ ($5d^4$) ions tend to be non-magnetic $J=0$, since orbital and spin moments cancel each other, see e.g.~\cite{streltsov2017}.  The non-zero magnetic moment on Ir$^{5+}$ ($5d^4$) is related to strong distortions of the IrO$_6$ octahedra \cite{Nag2016} and formation of molecular orbitals in Ir-Ir dimers \cite{streltsov2014a}.

{\bf S-type altermagnet: Ba$_3$NiRu$_2$O$_9$}. 
In contrast to the previously discussed M-type altermagnetism, the magnetic structure in this case is collinear, and altermagnetism manifests itself through spin splitting along certain high-symmetry directions in the electronic band structure. Fig.~\ref{Ba3NiRu2O9-ek}(a) illustrates $\varepsilon(\vec{k})$ and one can readily observe such splitting along the $M$-$A$ direction. There can be no linear Hall effect (magneto-optics) in S-type altermagnets, but one would expect a piezomagnetic effect.

The direct piezomagnetic effect results in onset of magnetization $m_{i} = \Lambda_{ijk} \sigma_{jk}$ as a response to stress given by $\sigma_{jk}$ tensor ($i,j,k$ are Cartesian coordinates). There is only one non-zero parameter of the piezomagnetic tensor $\Lambda_{ijk}$: $\Lambda_{112}=\Lambda_{211} = -2\Lambda$, $\Lambda_{222} = \Lambda$ for $6'/m'mm'$ MPG of Ba$_3$NiRu$_2$O$_9$. This means that distortion in the $x$ direction can generate transverse magnetization (here and below we have in mind the total magnetization per u.c.) along $y$ ($\Lambda_{211}$), strain along $y$ can result in non-zero $m_y$ ($\Lambda_{222}$), while other components must vanish (for comparison, one might expect the longitudinal effect: $\sigma_{xx} \to m_x$, $\sigma_{yy} \to m_y$, and $\sigma_{zz} \to m_z$ in Ba$_3$CoIr$_2$O$_9$).

\begin{figure}[t!]
  \includegraphics[width=0.49\textwidth]{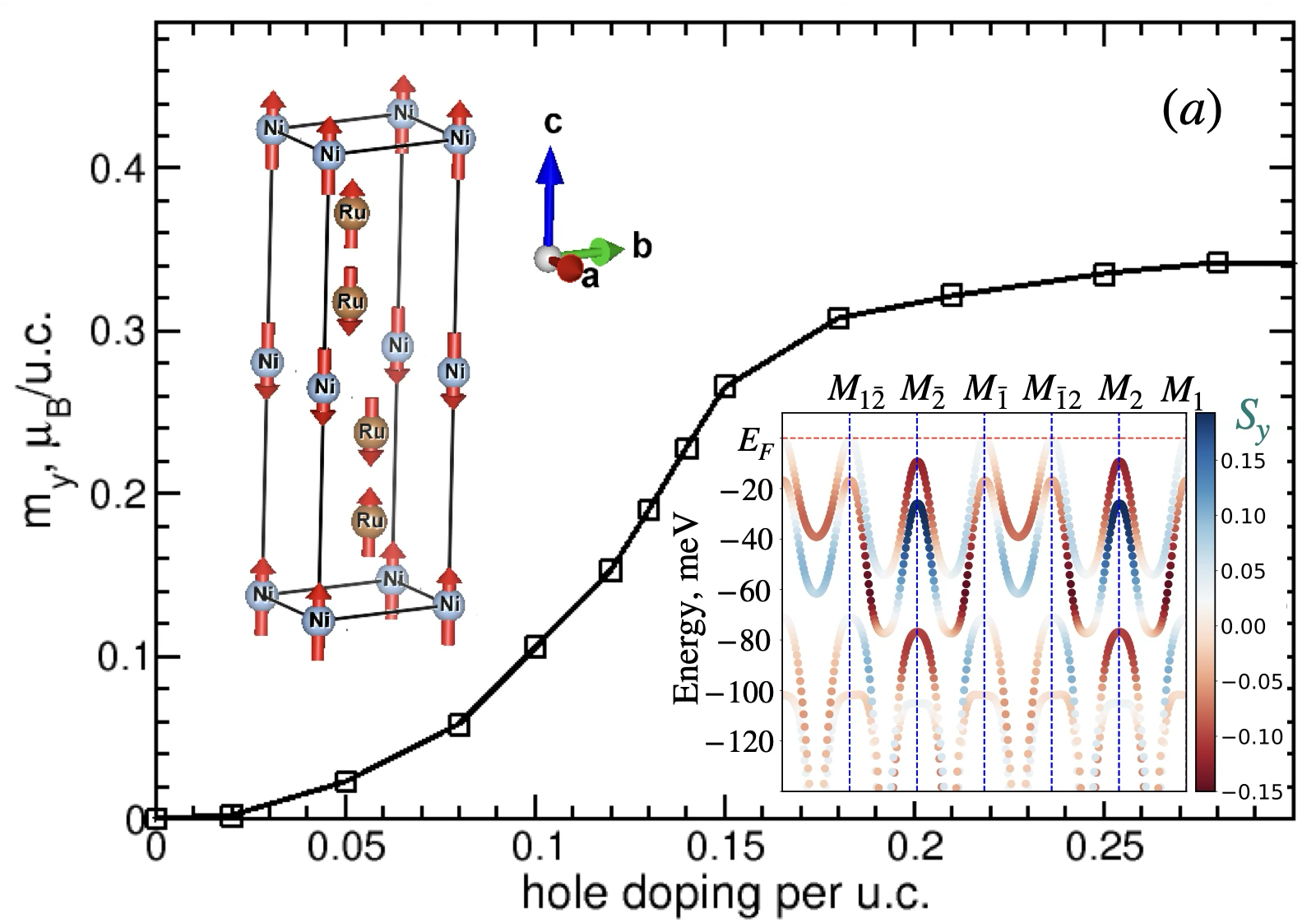}
  \includegraphics[width=0.49\textwidth]{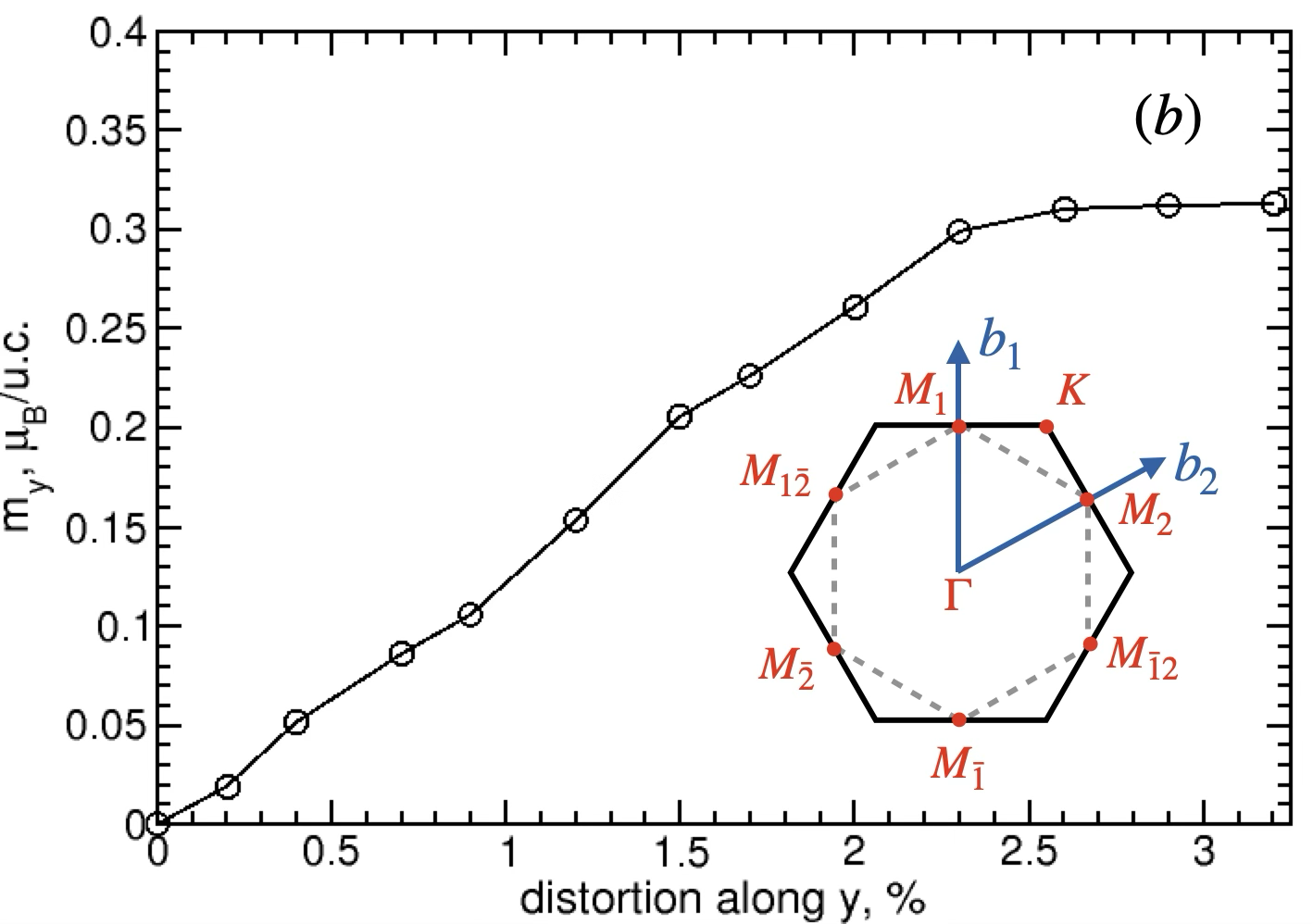}
\caption{\label{MvsDelta} Illustration of the piezomagnetic effect. (a): dependence of magnetization on doping in Ba$_{3-\delta}$NiRu$_2$O$_9$ as obtained in DFT+U+SOC calculations with 1\% tensile strain along $y$ direction applied. (b): magnetization as a function of distortion for hole doping 0.1 per unit cell (u.c.). Insets show: (a) left: experimental magnetic structure in u.c. containing 2 formula units (4 Ru and 2 Ni ions), the $c$-axis is along $z$; right: DFT+U+SOC band structure colored according to $S_y$ (for 1\% tensile strain). (b)  $q_z = 0$ cross-section of the Brillouin zone. }
\end{figure}

Naively, one would expect that the antisymmetric Dzyaloshinskii-Moriya interaction (activated by the lowering of symmetry from $P6_3/mmc$ to $Cmcm$ due to strain) can result in canting of the magnetic moments and the formation of a net ferromagnetic moment. However, direct DFT+U+SOC calculations, taking into account both possible non-collinearity and spin-orbit coupling, show a vanishingly small net magnetization of $\sim 10^{-3} \, \mu_B / 2 \, \text{f.u.}$, even when 1\% strain along the $y$-axis is applied. This can be attributed to a weaker exchange interaction; the N\'eel temperature in Ba$_3$NiRu$_2$O$_9$ is two orders of magnitude lower than in Ba$_3$CoIr$_2$O$_9$, despite their similar ground-state magnetic configurations (see Table~\ref{6H-AM-table} and the insets to Figs.~\ref{Ba3CoIr2O9-sigma} and \ref{MvsDelta}).

Analysis of the strain effect on the electronic structure reveals two significant outcomes. First, it lifts the degeneracy for some of the $k$-points; for example, $M_1=(1/2,0,0)$ and $M_2=(0,1/2,0)$ become inequivalent, as shown in Fig.~\ref{Ba3NiRu2O9-ek}(b). Second, there is a splitting of bands with different spin characters. Consequently, doping Ba$_{3}$NiRu$_2$O$_9$ by holes is expected to result in an imbalance in the spin projections, leading to the formation of a net magnetization. This is precisely what direct DFT+U+SOC calculations demonstrate.

In Fig.~\ref{MvsDelta}(a), the dependence of the net magnetization along the $y$-axis on hole doping, $n_h$, at a fixed tensile strain of 1\% along the $y$-axis is shown. It should be noted that the doping leaves the total (per u.c.) $m_x$ and $m_z$ components vanishing, exactly following the symmetry prescriptions. Moreover, the hole doping dependence can be rationalized at a qualitative level if the band structure is colored according to spin projections.

In right inset of Fig.~\ref{MvsDelta}(a) the top of valence band (for 1\% strain) is plotted if one goes along path including all M-points (their positions in Brillouin zone are shown in the top inset). Gradual doping increase up to 0.02 holes per u.c. practically does not change $m_y$, since electronic bands have small and nearly symmetric projections of positive and negative $S_y$.  Corresponding depopulation of states close to $M_{1 \bar 2}$, $M_{\bar 1}$, $M_{\bar 1 2}$, and $M_{1}$ points results in small $S_y$. Further doping shifts the Fermi level on more than 10 meV and to start depopulate states in $M_2$ and $M_{\bar 2}$ points with much larger by absolute value $S_y$. As a result magnetization along $y$ grows rapidly. Increasing doping further on we finally reach bands lying deeper in energy with opposite sign of $S_y$ and $m_y (n_h)$ saturates. The same band structure is symmetric with respect to $S_x$, as shown in Fig.~S1 in the supplemental materials~\cite{supp}. This explains why $m_x$ remains vanishing.

Doping can be achieved through small Ba non-stoichiometry. For instance, with $\delta = 0.1$ in Ba$_{3-\delta}$NiRu$_2$O$_9$, we obtain 0.2 holes per formula unit, or $n_h = 0.4$ holes per u.c. However, experimentally, it is more practical to apply external strain. The corresponding variation of induced magnetization is shown in Fig.~\ref{MvsDelta}(b) for a doping level of $n_h = 0.1$ holes per u.c. ($\delta = 0.05$).

There are two antiferromagnetic sub-lattices (Ni and Ru) in Ba$_3$NiRu$_2$O$_9$, and it is interesting to study which one contributes the most to the emerging magnetization. For example, at a doping level corresponding to 0.35 holes per u.c., each Ru ion contributes $0.1\,\mu_B$ to $m_y$ (i.e., $0.4\,\mu_B$/u.c.), while Ni induces a small correction in the opposite direction, $-0.03\,\mu_B$. This result is consistent with the band structure of Ba$_{3}$NiRu$_2$O$_9$: the Ru $t_{2g}$ states form the top of the valence band, and the spin splitting discussed above occurs precisely in these bands. Moreover, the AFM Ru dimers are isolated in the structure of 6H perovskites, and the corresponding $d$ bands are relatively flat. As a result, the effect of spin splitting is pronounced, leading to a large piezomagnetic effect. Additionally, Ru$^{5+}$ ions with a $t_{2g}^3$ electronic configuration have the maximal spin moment $S=3/2$ (the large $t_{2g}$-$e_{g}$ crystal-field splitting in $4d$ transition metals breaks Hund's rules, resulting in a situation where the $t_{2g}$ sub-shell is completely occupied first; see, e.g., \cite{khomskii2024}).

\hfill \break {\bf \noindent DISCUSSION AND CONCLUSIONS\\}
\begin{figure}[b!]
\includegraphics[width=0.49\textwidth]{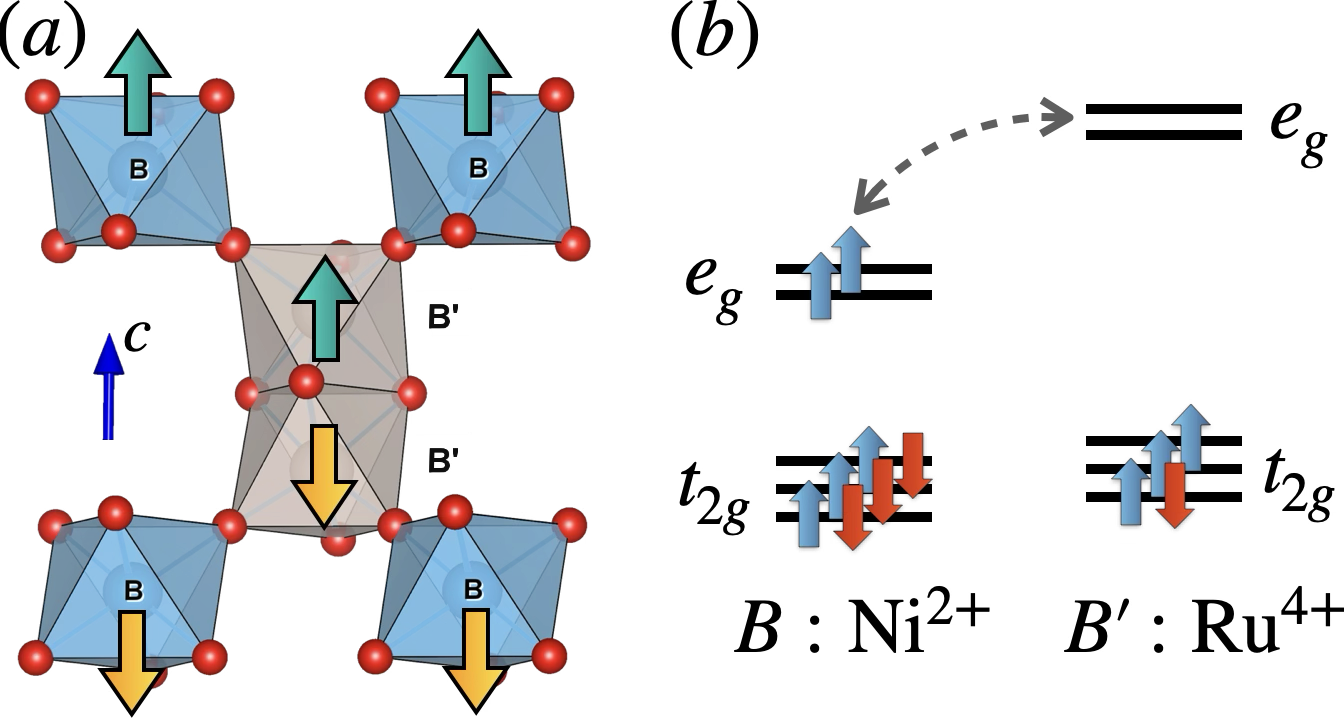}
\caption{\label{AFM-scheme} Sketches illustrating origin of antiferromagnetic interplane order for $B$ ions, due to ferromagnetic $B-B'$ and antiferromagnetic intradimer $B'-B'$ interactions (on example of Ni$^{2+}$ and Ru$^{4+}$ ions).}
\end{figure}

Altermagnetism is closely related to the crystal structure of 6H perovskites, which lack an inversion center but possess $C_2$ rotation axes that transform magnetic ions in different $B'$ planes into one another. Another crucial aspect is the nature of the exchange interaction. The BO$_6$ and B'O$_6$ octahedra share a common corner, and the $\angle$B-O-B' is close to 180$^{\circ}$, as shown in Fig.~\ref{AFM-scheme}(a). In this configuration, the exchange interaction between the $e_g$ orbitals is expected to be the strongest \cite{streltsov2017}. The $B'$ sites are occupied by $4d$ and $5d$ elements, which typically adopt a low-spin configuration with empty $e_g$ orbitals. The exchange interaction between empty and half-filled $e_g$ orbitals must be ferromagnetic, as illustrated in Fig.~\ref{AFM-scheme}(b). This is precisely the case in both Ba$_3$NiRu$_2$O$_9$ and Ba$_3$CoIr$_2$O$_9$: the $3d$ metals occupying $B$ sites order ferromagnetically with respect to its nearest-neighbor $4d/5d$ elements and, more importantly, this results in ferromagnetism in the $ab$ planes, as shown in the insets of Figs.~\ref{Ba3CoIr2O9-sigma}(a) and \ref{MvsDelta}(a). The exchange interaction between the $t_{2g}$ orbitals in $B'$ dimers with a common face tends to be antiferromagnetic, which, together with the ferromagnetic $B$-$B'$ exchange, ultimately results in antiferromagnetic order between the two (ferromagnetic) $B'$ planes and altermagnetism, when we take into account the structural features of 6H perovskites (i.e., the $C_2$ axes).

If a non-magnetic ion such as Sb$^{5+}$ or Nb$^{5+}$ is placed into the $B'$ sites, the in-plane magnetic order between the $3d$ metals switches to antiferromagnetic. These materials either form a $120^{\circ}$ antiferromagnetic structure or have been suggested to exhibit spin-liquid behavior, as shown in Tables S1 and S3. In the ordered case, where the magnetic unit cell is larger than the crystallographic unit cell, these materials cannot formally be considered altermagnets. However, they might possess a ferromagnetic point group. Examples include Ba$_3$CoSb$_2$O$_9$ with the $\bar{6}m2$ MPG, Ba$_3$MnNb$_2$O$_9$ with the $3m$ MPG, and Ba$_3$MnSb$_2$O$_9$ with the $2$ MPG. 
This ``quasialtermagnetism'' with $\vec{Q} \ne 0$ is likely related to magneto-elastic interactions and is expected to be weaker than in conventional altermagnets. Nevertheless, these materials could still be of significant interest due to their potential for non-trivial magneto-optical or piezomagnetic responses, as well as other effects dictated by symmetry.

Thus, 6H perovskites based on $3d$ and $4d/5d$ magnetic ions represent a promising combination of crystal structure and exchange interactions for realizing altermagnetism. Through direct {\it ab initio} calculations, we not only demonstrate that they are indeed altermagnets but also reveal giant piezomagnetism in Ba$_{3}$NiRu$_2$O$_9$, which is an order of magnitude larger than previously expected~\cite{ma2021}. Further studies of other mixed perovskites with 4H, 9R, and other layer sequences may uncover additional altermagnets with extraordinary physical properties.

\hfill \break {\bf \noindent METHODS \\}
DFT calculations were performed by the Vienna ab initio simulation package (VASP)\cite{Kresse1996}. The Perdew-Burke-Ernzerhof version of exchange correlation potential was used~\cite{Perdew1997}. The crystal structures were taken from Refs.~\cite{Lightfoot1990} and \cite{doi2004}. Strong Coulomb correlations were treated by DFT+U approach\cite{Dudarev1998} with $U_{\rm Ni}=8$ eV, $U_{\rm Co}=7$ eV, $U_{\rm Ru}=3$ eV, $U_{\rm Ir}=1.5$ eV, $J_{\rm Ni}=J_{\rm Co}=1$ eV, $J_{\rm Ru}=0.7$ eV and $J_{\rm Ir}=0.5$ eV, close to typical values previously used for transition metal oxides~\cite{streltsov2013,muthuselvam2014,terzic2015,zvereva2016}. For both materials experimental magnetic structure was used~\cite{Lightfoot1990,doi2004}. The $6 \times 6 \times 4$ mesh in the Brillouin zone was utilized for self-consistent calculations (increase up to $7 \times 7 \times 5$ k-points did not change the results). The calculations were performed using the tetrahedron method with Bl\"ochl corrections, except for magneto-optics, where the Gaussian broadening \cite{methfessel1989} of 50 meV was chosen (complex shift in the Kramers-Kronig transformation was set to 100 meV in this case).  The convergence criteria for electronic iterations was set to $2 \times 10^{-7}$ eV.  For piezomagnetic effect calculations the spin-orbit coupling was take into account. We used experimental crystal structure in these calculations, strains were introduced via modification of the unit cell, atomic positions were not relaxed.

\hfill \break {\bf \noindent DATA AVAILABILITY\\} 
All data generated or analyzed during this study are available from the
corresponding authors upon reasonable request.

\hfill \break {\bf \noindent ACKNOWLEDGMENTS\\} 
S.V.S. would like to thank M. Braden, V. Mineev  for stimulating discussions and A. Gubkin for helping with symmetry analysis.

The first-principle calculations were supported by the Russian Science Foundation via project RSF 23-12-00159, while symmetry analysis was performed with support by Ministry of Science and Higher Education of the Russian Federation. S.-W.C. was supported by the W. M. Keck foundation grant to the Keck Center for Quantum Magnetism at Rutgers University.

\hfill \break {\bf \noindent AUTHOR CONTRIBUTIONS\\} 
S.-W.C. and S.V.S. conceived the project. S.V.S. performed DFT calculations and symmetry analysis. S.-W.C. and S.V.S. discussed all results. S.V.S. wrote the paper. 

\hfill \break {\bf \noindent  COMPETING INTERESTS\\} 
The authors declare no competing interests.

\hfill \break {\bf \noindent  ADDITIONAL INFORMATION\\} 
\noindent {\bf Supplementary information} The online version contains supplementary material
available at to-be-add-by-publisher.

\noindent {\bf Correspondence} and requests for materials should be addressed to S. V. Streltsov.

\bibliographystyle{naturemag}
\bibliography{ref}

\end{document}